\documentstyle[sprocl,epsf]{article}

\newfont{\bg}{cmr10 scaled\magstep4}

\newcommand{\bigzerou}{\smash{\lower1.ex\hbox{\bg 0}}}

\newcommand{\BN}{\begin{eqnarray}}
\newcommand{\EN}{\end{eqnarray}}
\newcommand{\NN}{\nonumber}
\begin{document}
\vspace*{-2cm}
\begin{flushright}
KANAZAWA 96-08, June, 1996
\end{flushright}
\vspace{1cm}
\title{
Dual Lattice 
Blockspin 
Transformation and 
Monopole 
condensation  in QCD
\footnote{
This is based on works done in collaboration with Y.Matsubara (Nanao), 
S.Kitahara (Jumonji), H.Shiba (RCNP), S.Ejiri, S.Okude, 
F.Shoji, N.Nakamura, H.Kodama, M.Sei,
V.Bornyakov (IHEP).
}
}
\author{
 Tsuneo Suzuki
\footnote{ E-mail address: suzuki@hep.s.kanazawa-u.ac.jp}
}

\address{
{\em Department of Physics, Kanazawa University, Kanazawa 920-11, Japan
}}

\maketitle\abstracts{
Recent studies of confinement based on the idea of abelian 
monopole condensation are reviewed briefly. 
Emphasis is placed on the approach 
to get the effective monoole action using the blockspin transformation on 
the dual lattice. 
The trajectory obtained looks to be the renormalized one
in $SU(2)$ QCD.
A disorder parameter of confinement is constructed.
Monopole condensation occurs also in $SU(3)$ QCD.
}

\section{Introduction}
It is crucial to understand the mechanism of quark confinement 
in order to explain hadron physics out of QCD.
The 'tHooft idea of abelian projection 
of QCD \cite{thooft} is very attractive.
The abelian projection is to fix the gauge in such a way that the maximal 
torus group remains unbroken.
After the abelian projection, monopoles appear as a topological quantity 
in the residual abelian channel.
QCD is reduced to an abelian theory with electric charges and monopoles.
If the monopoles make Bose condensation, charged quarks and gluons 
are confined due to the dual Meissner effect. 

Based on this standpoint, we  have studied quark
confinement mechanism and hadron physics performing
Monte Carlo simulations of abelian projection 
in lattice QCD
\cite{suzu90,hio91a,hio91b,hio92,suzu93,kita93,matsu94,shiba94a,shiba94b,ejiri95a,shiba95a,kita95,suzu95a,suzu95b,suzu95c,ejiri95b,ejiri95c,suzurev}.
The aim of the study is to ascetain correctness of 
the picture, that is, to check if monopole condensation really 
occurs in QCD.
Here I  review the results compactly.\footnote{ 
We have also studied hadron physics based on 
an infrared effective Lagrangian constructed 
directly from QCD on the assumption of the above picture
\cite{suzu88,maedan89,maedan90,monden,kamizawa,matsurev}. 
}

\section{Abelian dominance and monopole dominance}
Our procedure is as follows:
\begin{enumerate}
\item
Vacuum configurations of link variables $\{U(s,\mu)\}$ are generated 
 with the Wilson action. 
\newpage
\vspace*{-1.cm}
\epsfxsize=5.cm
\begin{flushleft}
\leavevmode
\epsfbox{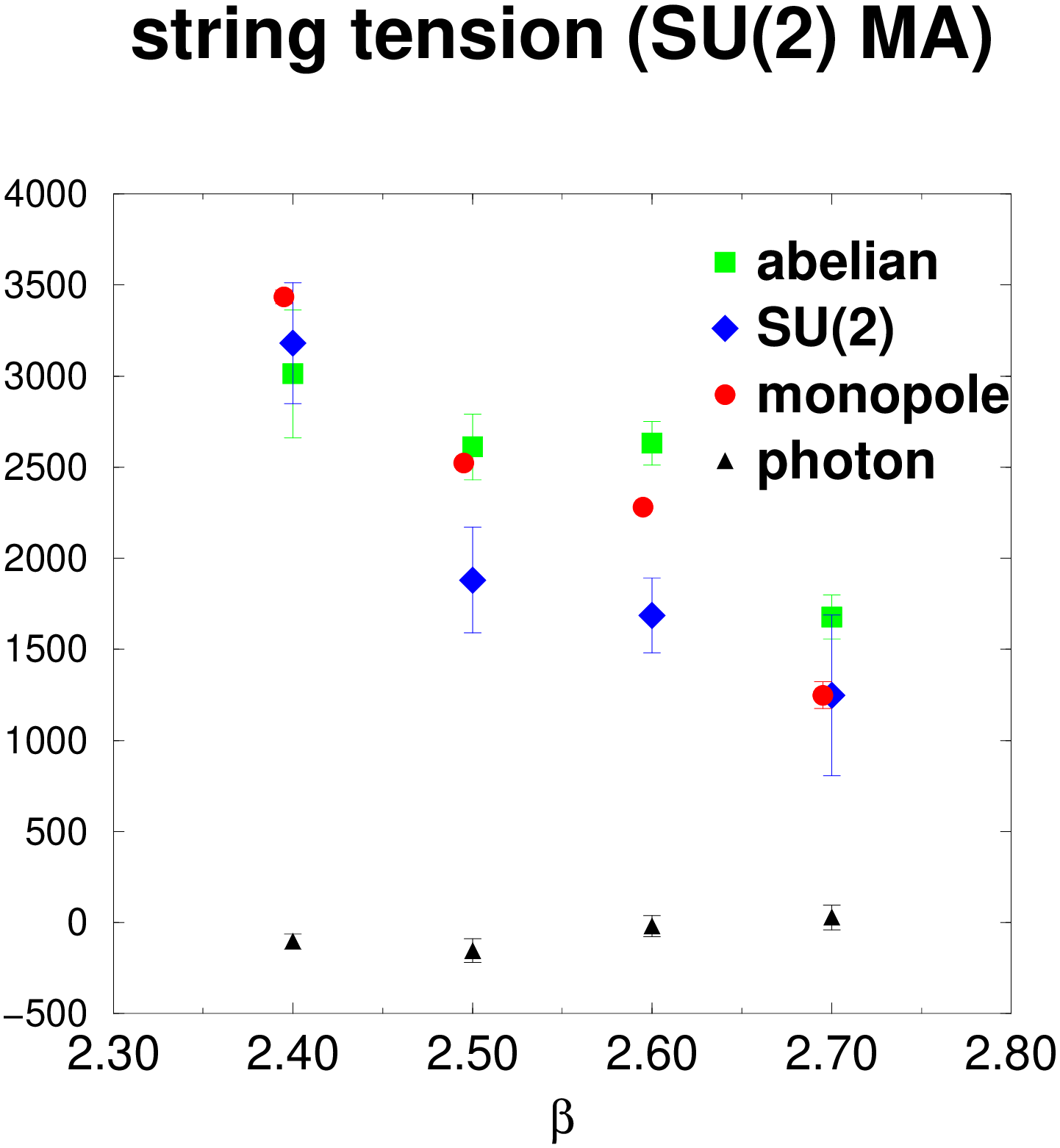}
\vspace{-1cm}
\end{flushleft}
\vspace{-2cm}
\parbox{5.cm}{
\begin{flushleft}
Fig.1
Monopole and photon 
contributions to the string 
tension in MA gauge in $SU(2)$
QCD.
\end{flushleft}
}
\vspace{-8.cm}
\begin{flushright}
\epsfxsize=5.cm
\leavevmode
\epsfbox{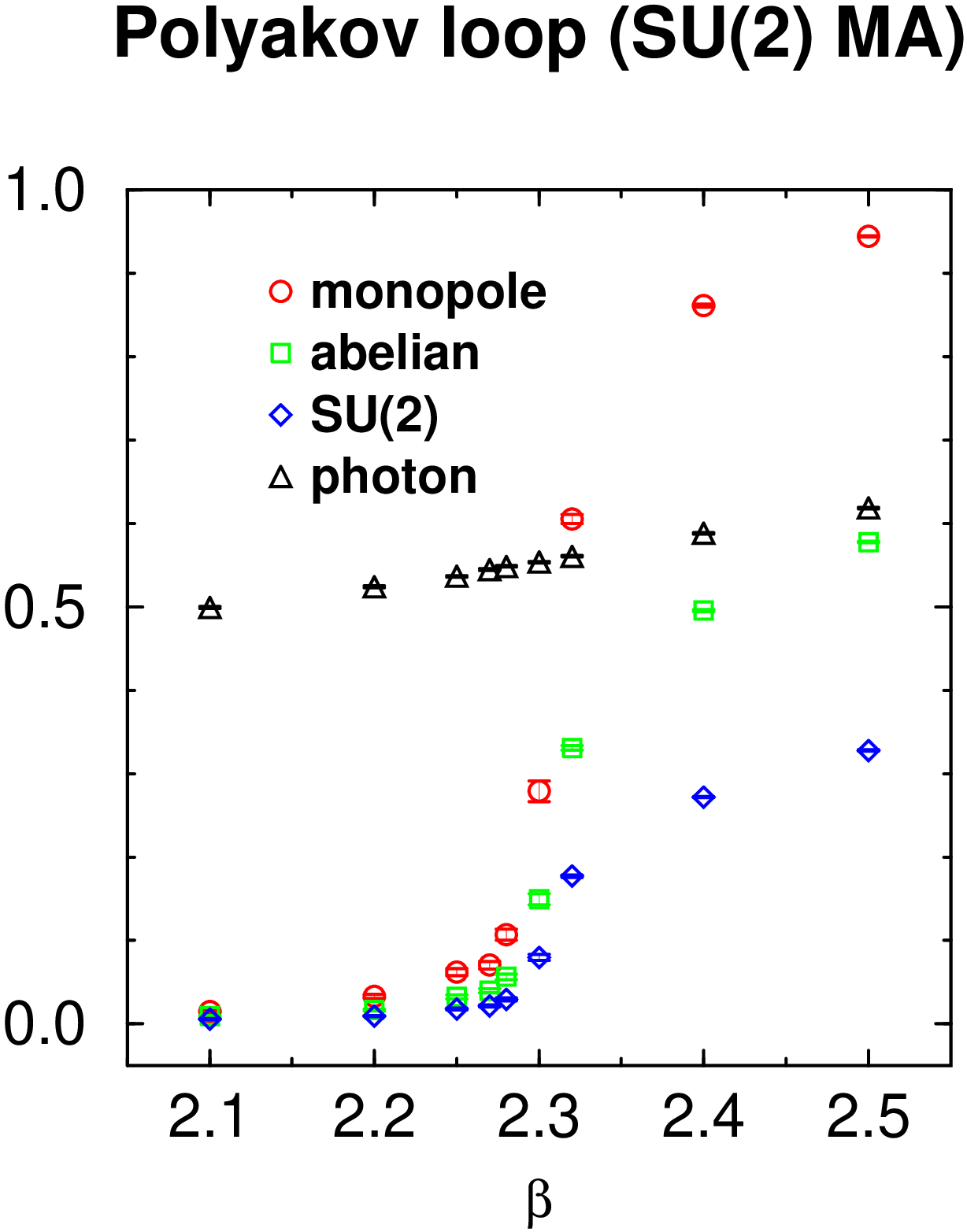}
\end{flushright}
\vspace{-1.5cm}
\begin{flushright}
\parbox{5.cm}{
\begin{flushleft}
Fig.2 Polyakov loops in MA gauge.
\end{flushleft}
}
\end{flushright}

\vspace{1.cm}
\begin{center}
\epsfxsize=6cm
\leavevmode
\epsfbox{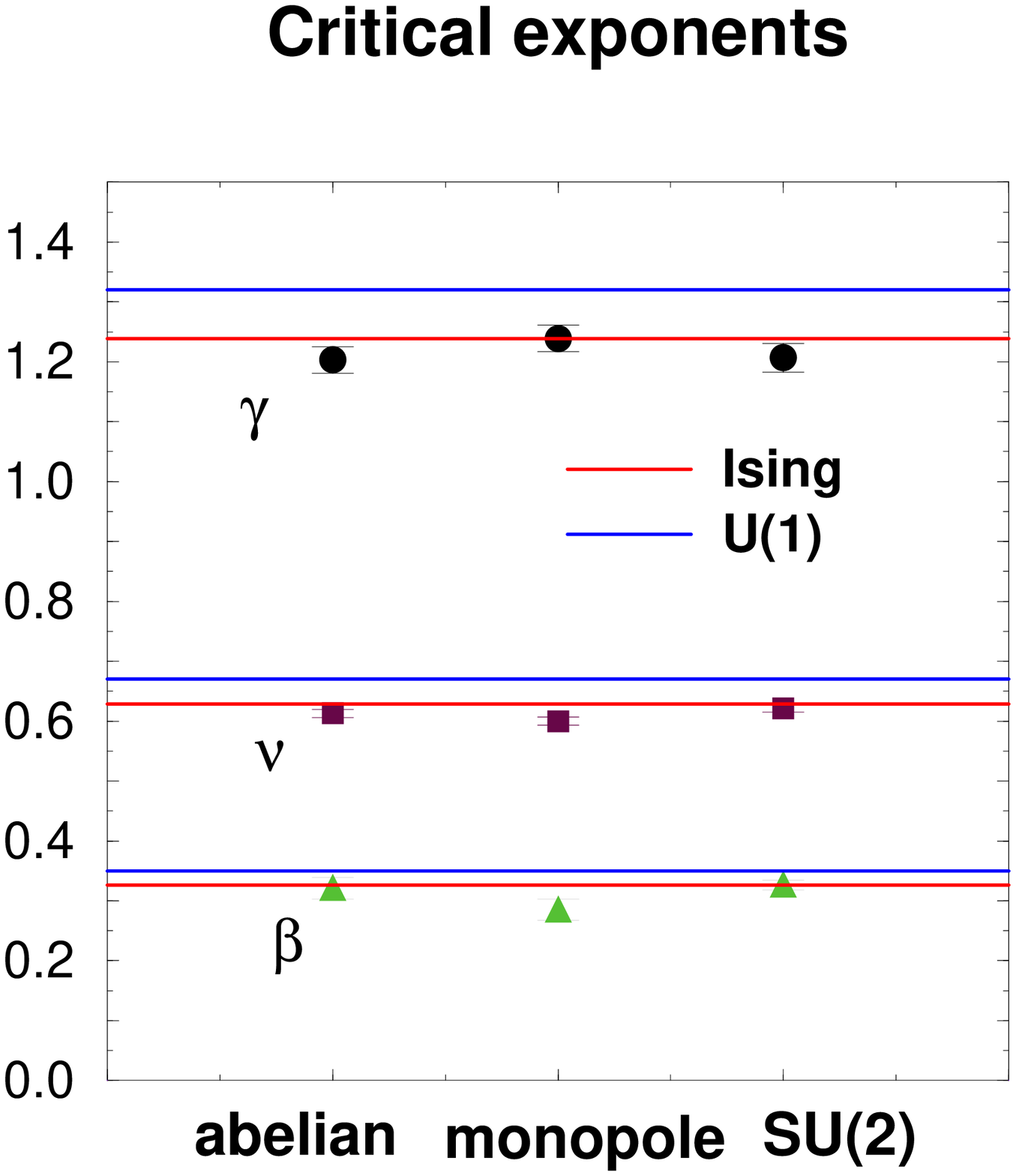}
\end{center}
\vspace{-2.5cm}
\begin{center}
\parbox{6cm}{
\begin{flushleft}
Fig.3 Critical exponents in SU(2).
\end{flushleft}
}
\end{center}

\newpage
\item
We perform the abelian projection. One of interesting gauges is the maximally 
abelian (MA) gauge \cite{kron}.  
Define a matrix in $SU(2)$ QCD
\begin{eqnarray}
X(s) & = & \sum_{\mu}[
           U(s,\mu)\sigma_3 U^{\dagger}(s,\mu) 
      + U^{\dagger}(s-\hat{\mu},\mu)\sigma_3 U(s-\hat{\mu},\mu)] \\
     & = & X_1 (s)\sigma_1 + X_2 (s)\sigma_2 + X_3 (s)\sigma_3.
\end{eqnarray}
Then a gauge satisfying $X_1 (s)=X_2 (s)=0$ is the MA gauge.
\item
Separate abelian link variables 
$u(s,\mu)$ as 
$V(s)U(s,\mu)V^{\dagger}(s+\hat{\mu})=C(s,\mu)u(s,\mu)$
 to obtain an ensemble of $\{u(s,\mu)\}$. 
\item
We construct monopole currents $k_{\mu}(s)$ following the 
DeGrand-Toussaint method \cite{degrand}.
\item
We measure expectation values of $U(1)\times U(1)$ invariant operators
$O(u(s,\mu))$ and operators composed of  monopole currents 
$O(k_{\mu}(s))$.
\end{enumerate}

We have found that 
important features of confinement, i.e., the string tension and the 
charateristic behaviors of the Polyakov loops are well reproduced in 
terms of the abelian operators $O(u(s,\mu))$ and also $O(k_{\mu}(s))$
in MA gauge \cite{suzu90,suzu93,stack,shiba94b,suzurev}.
Abelian and monopole Polyakov loops are 
different operators, but they
give almost equal critical exponents \cite{kita_lat96}.
See Figs.1$\sim$3.

\section{Monopole  action and condensation}
The above abelian dominance suggests that a set of $U(1)^2$ 
invariant operators 
$\{O(u(s,\mu))\}$ are enough to describe confinement. 
Then there must exists an effective abelian action 
$S_{eff}(u)$ describing confinement.
We tried to derive $S_{eff}(u)$ using Schwinger-Dyson equations, but failed 
to get it in a compact and local form \cite{suzu93}.
$S_{eff}(u)$ contains larger and larger loops as $\beta$.

Shiba and Suzuki \cite{shiba94a,shiba95a} tried to 
perform a dual transformation of $S_{eff}(u)$ in $SU(2)$ QCD and to obtain 
the effective $U(1)$ action in terms of monopole currents, 
extending the Swendsen method \cite{swendsn}.
To study the long range behavior is important in QCD, they have considered also
extended monopoles \cite{ivanenko}. 
The extended monopole currents are defined by the number of the Dirac strings
 surrounding an extended cube:
\begin{eqnarray}
k_{\mu}^{(n)}(s) 
    & = & \sum_{i,j,l=0}^{n-1}k_{\mu}(ns+(n-1)\hat{\mu}+i\hat{\nu}
     +j\hat{\rho}+l\hat{\sigma}),
\end{eqnarray}
where $k_{\mu}(s)$ is the ordinary monopole current \cite{degrand}.
Considering extended monopoles corresponds to performing a block spin 
transformation on a dual lattice \cite{shiba94a} and so it is 
suitable for exploring the long range property of QCD.

How about the case of SU(3) QCD? 
There are two independent (three with one constraint 
$\sum_{i=1}^{3}k^i_{\mu}(s)=0$) currents.
When considering the two independent currents, their entropies are 
difficult to evaluate. Hence we try to evaluate the effective 
monopole action, paying attention to only one monopole current. Then 
we can get the monopole action similarly as in SU(2).

The partition function of interacting monopole currents is expressed as
\begin{eqnarray}
  Z=(\prod_{s,\mu} \sum_{k_{\mu} (s)=-\infty}^\infty ) \,
    (\prod_{s} \delta_{\partial '_{\mu}k_{\mu}(s) ,0 } ) \,
    \exp (-S[k]) .
\label{eqn:pfunm}
\end{eqnarray}
It is natural to assume $S[k] = \sum_i f_i S_i [k]$. Here
$f_i$ is a coupling constant of an interaction $S_i [k]$. 
For example, $f_1$ is the coupling of the self energy term  
$\sum_{n,\mu}(k_\mu(s))^2$, $f_2$ is the coupling 
of a nearest-neighbor interaction term
 $\sum_{n,\mu} k_\mu(s) k_\mu(s+\hat\mu)$ 
and $f_3$ is the coupling of another nearest-neighbor term 
$\sum_{n,\mu\neq\nu} k_\mu(s) k_\mu(s+\hat\nu)$.

The monopole actions are obtained locally enough 
for all extended monopoles considered even in the scaling region.
They are 
lattice volume independent.
The coupling constant $f_1$ of the self-energy term is dominant 
and the coupling constants decrease 
rapidly as the distance between the two monopole currents increases
as seen in Figs. 4 and 5.

To study monopole dynamics, we have also studied the length 
of monopole loops \cite{kita95}. We have found that
the value of the action is proportional to the length $L$ of the loop
and is well approximated by $f_1 \times L$. 

As done in compact QED \cite{bank}, 
the entropy of a monopole loop can be estimated as $\ln 7$ 
per unit loop length.
Since
the action is approximated by 
the self energy part $f_1 L$,
the free energy per unit monopole loop length is approximated by 
 $( f_1-\ln 7 ) $ .
If $f_1 < \ln 7$, the entropy dominates over the energy, 
which means condensation of monopoles.
In Figs.6 and 7, $f_1$ versus $\beta$ for various extended 
monopoles on $24^4$ 
lattice is shown in comparison with the entropy value $\ln 7$.
Monopole condensation occurs  
both in $SU(2)$ and $SU(3)$.
Each extended monopole has its own $\beta$ region where the condition
$f_1 < \ln 7$ is satisfied.
When the extendedness is bigger, larger $\beta$ is included in such a 
region. Larger extended monopoles are more important in determining 
the phase transition point.

The behaviors of $f_i$ are 
different for different extended 
monopoles. However, 
if we plot them versus 
$b=n\times a(\beta)$,
we get a unique curve as in Fig. 8. The 
coupling constants seem to depend only on $b$, not on the extendedness 
nor $\beta$.  
There is a critical $b_c$ corresponding to critical $\beta^n_c$,
 i.e., $b_c =na(\beta^n_c)$.
On the other hand,
the scaling is not yet good in $SU(3)$.

\vspace*{1cm}

\epsfxsize=5cm
\begin{flushleft}
\leavevmode
\epsfbox{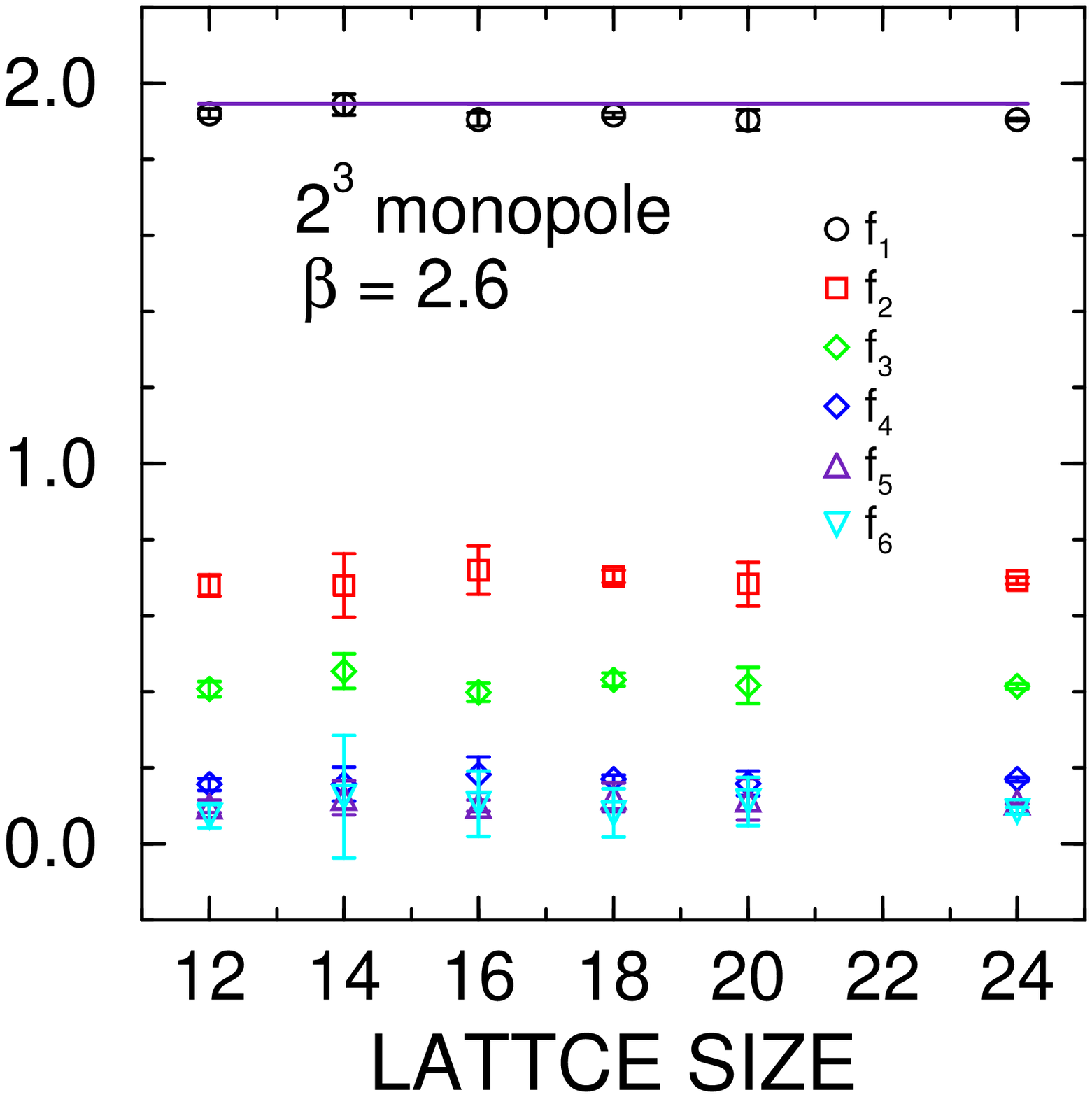}
\end{flushleft}
\vspace{-1.5cm}
\begin{flushleft}
Fig.4\ Monopole action in $SU(2)$.
\end{flushleft}

\vspace{-7.cm}

\epsfxsize=5.5cm
\begin{flushright}
\leavevmode
\epsfbox{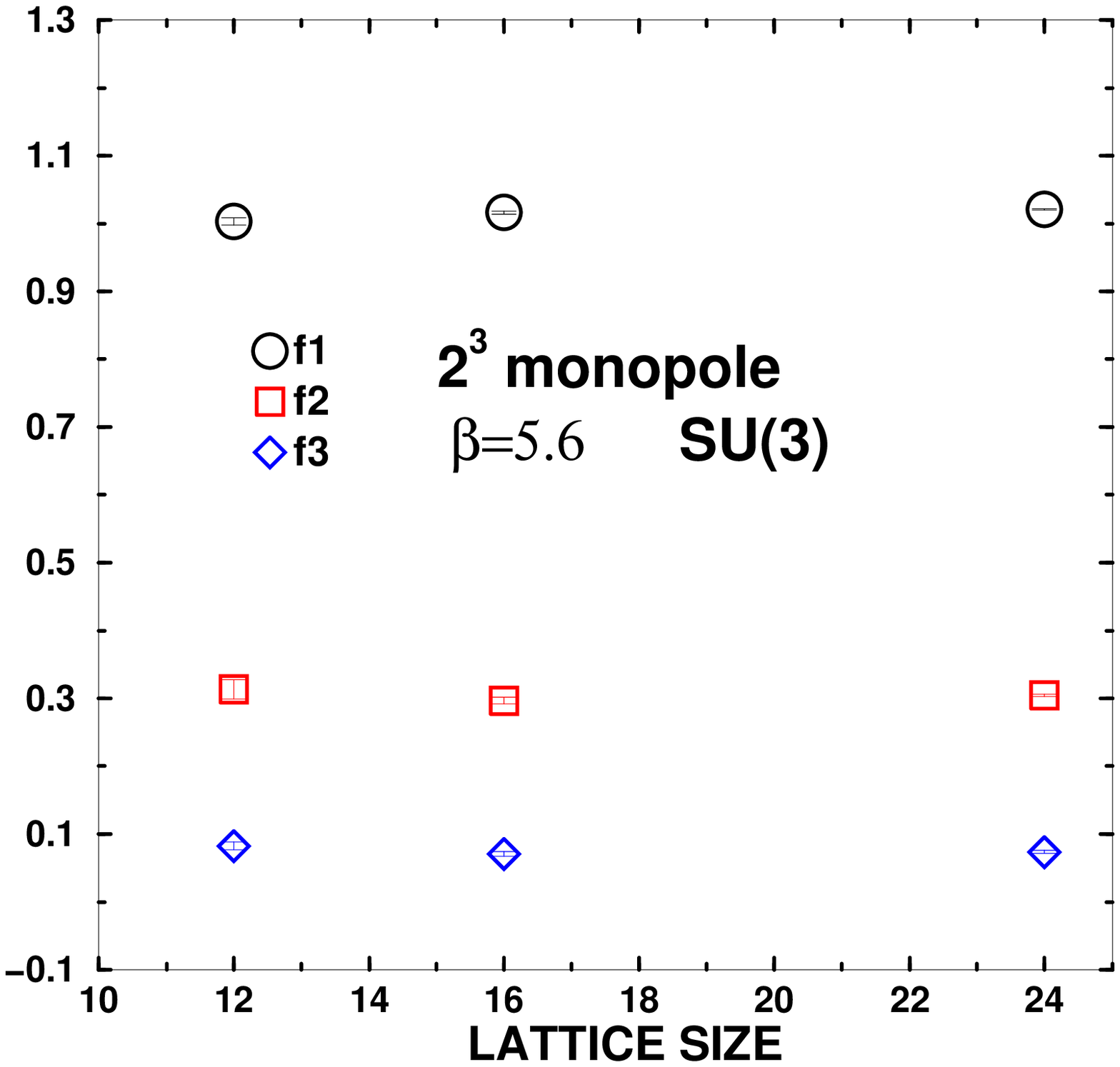}
\end{flushright}
\vspace{-2.cm}
\begin{flushright}
\parbox{5.5cm}{
\begin{flushleft}
Fig.5\ Monopole action in $SU(3)$.
\end{flushleft}
}
\end{flushright}

\vspace{1.5cm}

\epsfxsize=5.5cm
\begin{flushleft}
\leavevmode
\epsfbox{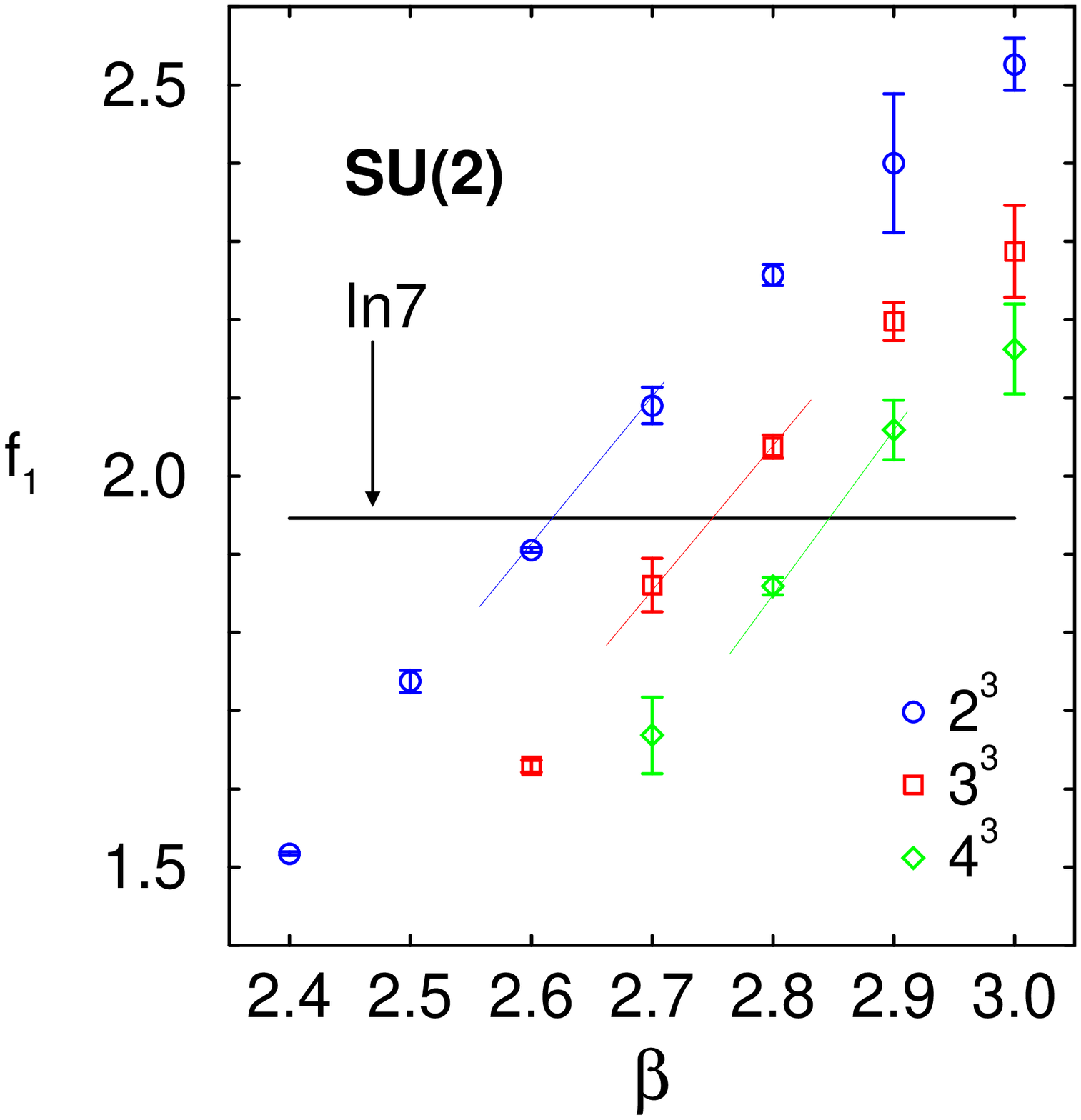}
\end{flushleft}
\vspace{-1cm}
\begin{flushleft}
Fig.6\ $f_1$ versus $\beta$ in SU(2).
\end{flushleft}

\vspace{-8cm}

\epsfxsize=5.5cm
\begin{flushright}
\leavevmode
\epsfbox{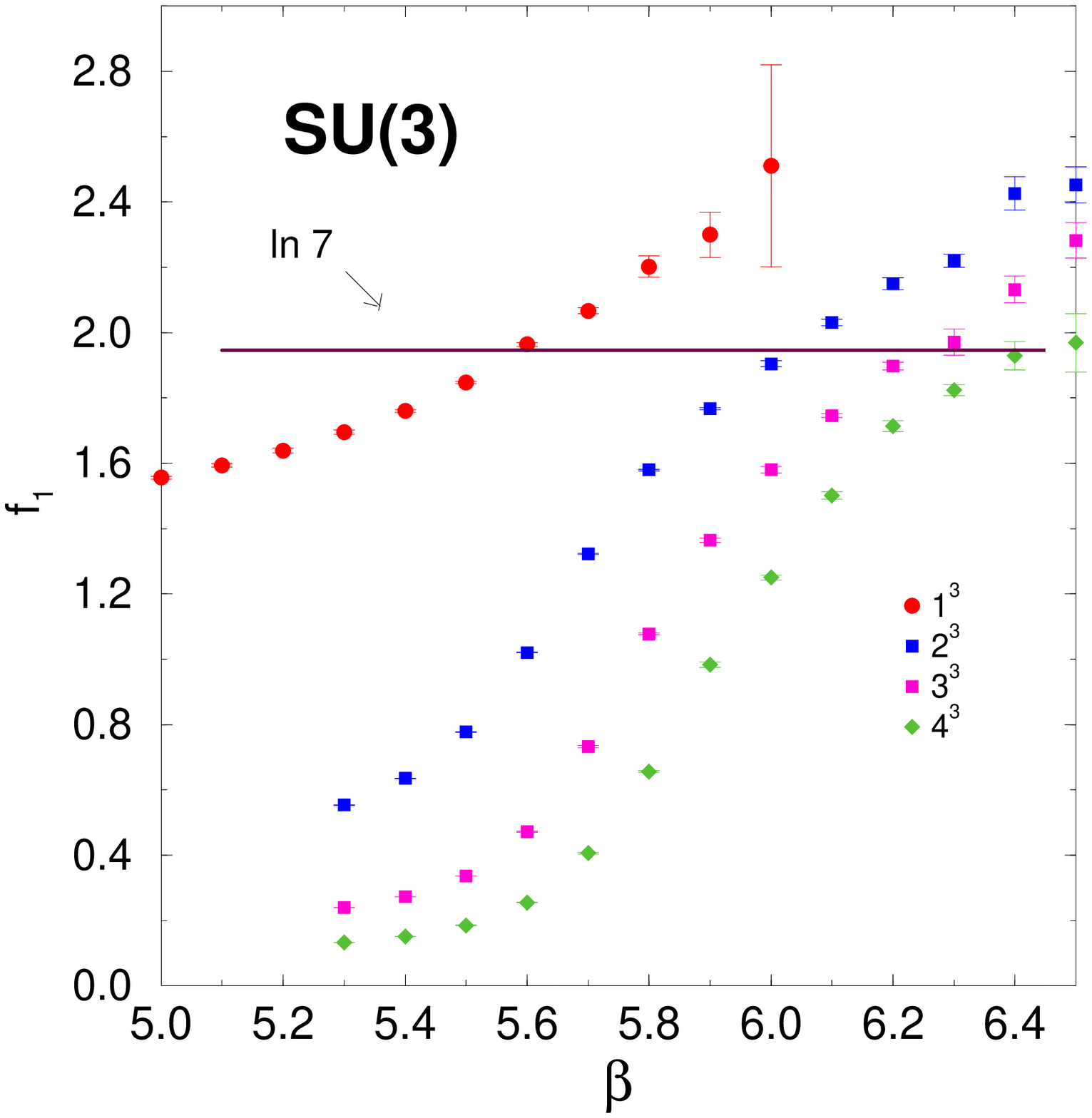}
\end{flushright}
\vspace{-1.cm}
\begin{flushright}
\parbox{5.5cm}{
\begin{flushleft}
Fig.7\ $f_1$ versus $\beta$ in SU(3).
\end{flushleft}
}
\end{flushright}

\newpage


\vspace*{-2.5cm}

\epsfxsize=5.5cm
\begin{flushleft}
\leavevmode
\epsfbox{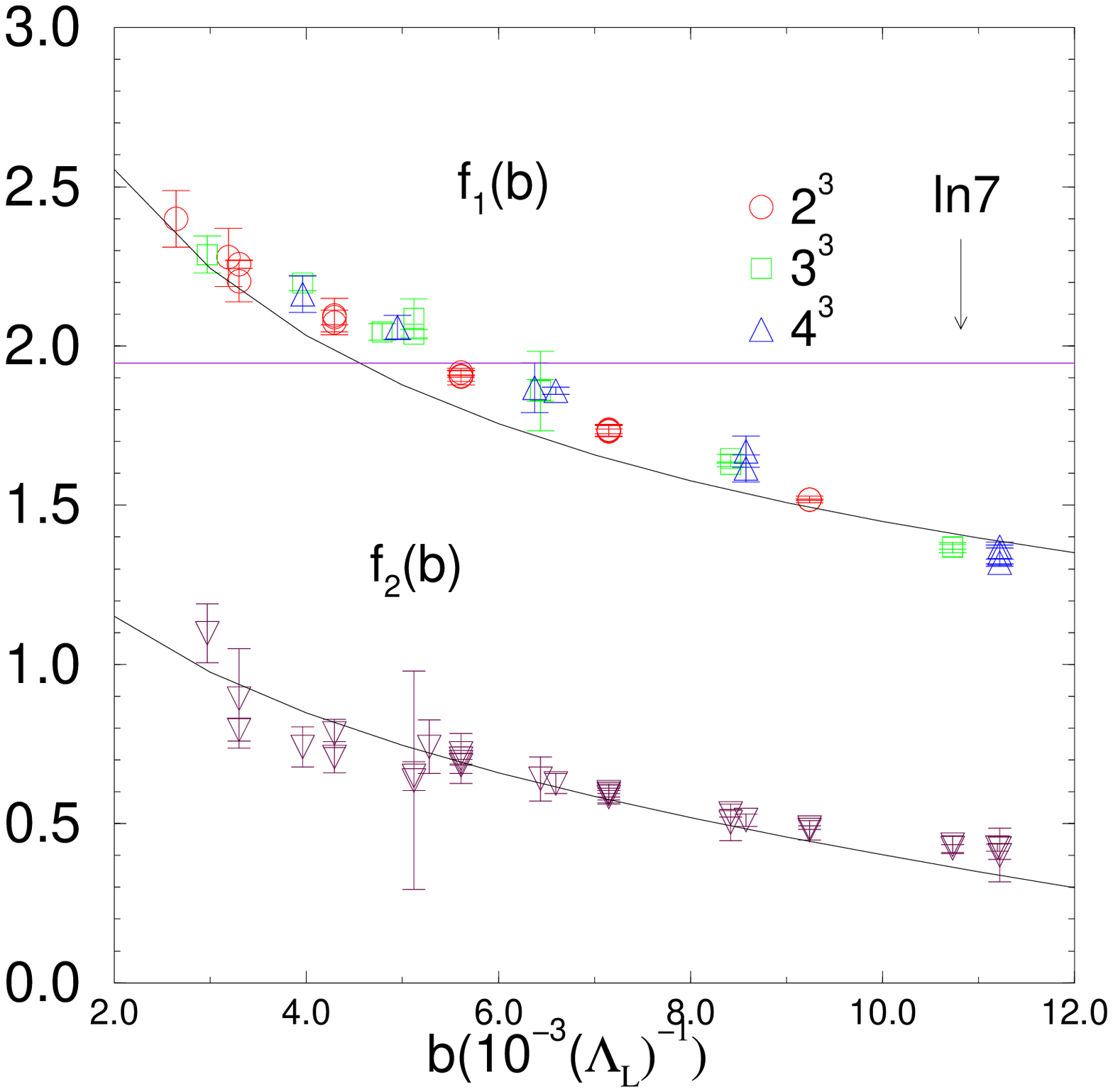}
\end{flushleft}
\vspace{-1cm}
\parbox{5.5cm}{
\begin{flushleft}
Fig.8\ Coupling constants $f_1$ \\
versus $b$ in SU(2).
\end{flushleft}
}

\vspace{-7.5cm}

\epsfxsize=5.5cm
\begin{flushright}
\leavevmode
\epsfbox{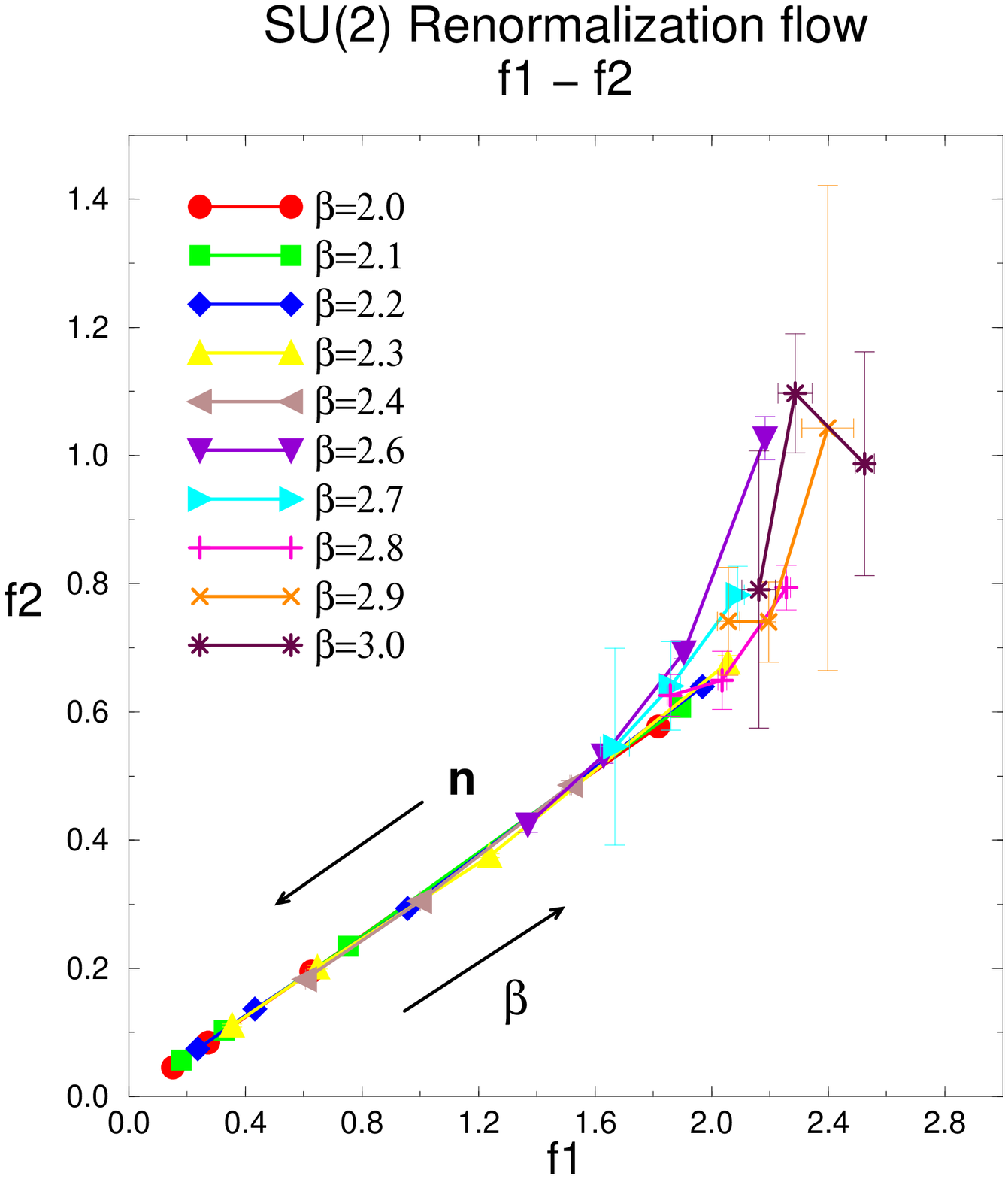}
\end{flushright}
\vspace{-1cm}
\begin{flushright}
\parbox{6cm}{
\begin{flushleft}
Fig.9\ Renormalization flow
 in SU(2).
\end{flushleft}
}
\end{flushright}

\vspace{1.5cm}

Now we can derive important conclusions at least in SU(2).
Suppose the effective monopole action remains the same for any 
extended
 monopoles in the infinite volume 
limit. Then the finiteness of $b_c =na(\beta^n_c)$ suggests 
$\beta^n_c$ becomes infinite when the extendedness $n$ 
goes to infinity. $SU(2)$ lattice QCD is always (for all $\beta$) 
in the monopole condensed and then in the quark confinement phase 
\cite{thooft}. This is one of what one wants to prove 
in the framework of lattice QCD.


Notice again that considering extended monopoles corresponds to performing 
a block spin transformation on the dual lattice. 
The above fact that the effective actions 
for all extended monopoles considered 
are the same for fixed $b$ means that the action may be  
the renormalized trajectory on which one can take the continuum limit.
See Fig. 9.
The explicit form of the obtained action will be discussed in 
elsewhere \cite{suzu_lat96}.


\section{Disorder parameter of confinement}
Since we have obtained the effective abelian U(1) action from SU(2) QCD,
it is possible to define a disorder parameter of confinement
\cite{giaco,misha} 
following the discussions done in compact QED \cite{froelich}
or in abelian Higgs model \cite{kennedy}.

Following Fr\"olich and Marchetti, we can define the correlation 
of two monopole operators using the differential form as 
\begin{eqnarray}
G_{\Psi}(x,y)= \frac{1}{Z}\sum_{^*k\in Z,\delta ^*k=0}
                \exp[-(^*k+^*\!\Psi-^*\!\omega,{\cal D}
                         (^*k+^*\!\Psi-^*\!\omega))],\NN
\end{eqnarray}
where 
$Z= \sum_{^*k\in Z,\delta ^*k=0}\exp[-(^*k,{\cal D}^*k)]$ is the 
partition function of the monopole current ensemble and $
^*\Psi=^*\!\Psi_3=d_3\Delta^{-1}_3(\delta_x-\delta_y)$ is the smeared string 
and the open current $^*\omega$ correnponding to the external 
monopole sources satisfies  
$\delta^*\omega=\delta_x-\delta_y$.
Then the disorder parameter can be defined as 
\BN
G_3=\mathop{\lim}_{\vert x-y\vert \to \infty}G_{\Psi}(x,y)\NN
\EN
\vspace{0.5cm}

Numerically, this shows a typical behavior of the disorder parameter 
of confinement as shown in Fig. 10.
Also 
one can define another but similar
form of the disorder parameter
following the discussions by  Kennedy and King \cite{kennedy}
in noncompact Abelian Higgs model. 
For details, see the reference \cite{nakamura}.

\vspace{1.5cm}

\epsfxsize=7cm
\begin{center}
\leavevmode
\epsfbox{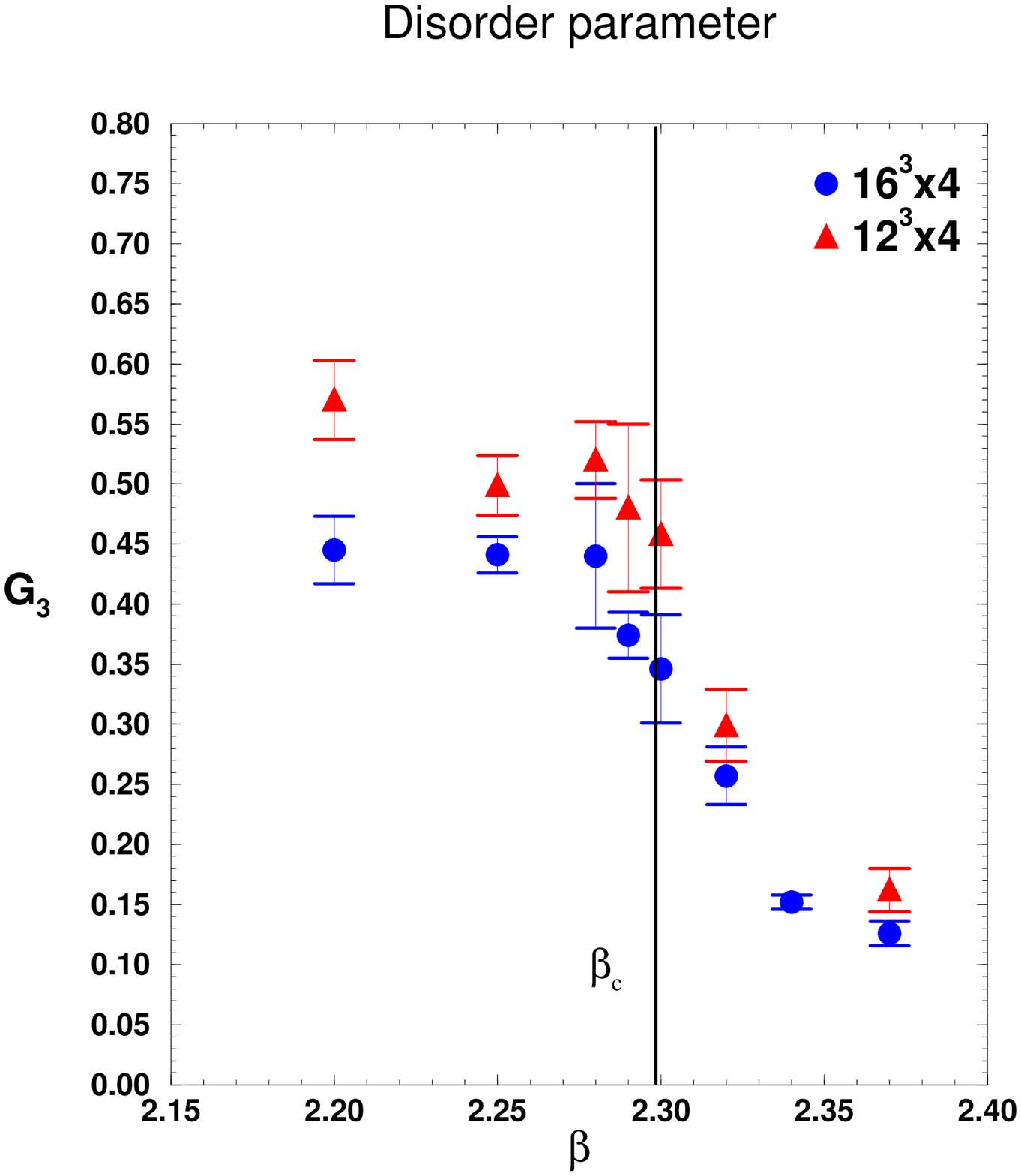}
\end{center}
\vspace{-1cm}
\begin{center}
\parbox{7cm}{
\begin{flushleft}
Fig.10\ Disorder parameter of confinement
 in SU(2).
\end{flushleft}
}
\end{center}


\section{Discussions and outlook}

A few comments are in order.
\begin{enumerate}
\item
Gauge independence should be proved if the monopole condensation 
is the real confinement mechanism.
My opinion is the following: 
gauge independent results will be obtained 
when we discuss long-range low-energy behaviors,
 if we go to the scaling region, i.e., for
large $\beta$ on larger lattice. 
Actually new interesting gauges have been found recently which show 
abelian and monopole dominances as in MA. See the reference \cite{suzu_lat96}.
\item
How to test the correctness of this idea?
The theory 
predicts the existence of 
an axial vector glueball-like state  C($J^{p}=1^{+}$)
and 
a scalar glueball-like state $\chi$($J^{p}=0^{+}$).
The masses seem to satisfy $m_c\sim m_{\chi}$ \cite{hay,hay93,matsu94}.
The masses could not be too heavy. They have to exist under 2GeV.
To evaluate the correlation between the state and the light hadrons
in MC simulations of full QCD is very important to derive the
total width and the branching ratios.
\end{enumerate}

\vspace{1.cm}

\section*{Acknowledgments}
The author is thankful to Y.Matsubara
for  collaboration and fruitful discussions.
This work is financially supported by JSPS Grant-in Aid for 
Scientific  Research (B) (No.06452028).

\end{document}